\documentclass[12pt]{article}
\usepackage{amsmath, amssymb, amsthm, fullpage, parskip, setspace}
\usepackage{natbib}
\usepackage{algpseudocode}
\usepackage{graphics, graphicx}
\usepackage{epsf}
\usepackage{float, multirow,parskip, subcaption, setspace}
\usepackage{comment}
\usepackage{url}
\usepackage{enumitem}
\usepackage{tikz}
\usepackage{bm, bbm}
\usetikzlibrary{shapes.geometric, positioning}
\usetikzlibrary{quotes, angles}
\usepackage{rotating}
\usepackage{xr, xr-hyper}
\usepackage{hyperref}
\usepackage{placeins}
\usepackage{algorithm2e}
\usepackage{xspace}
\usepackage{xcolor,soul}
\definecolor{vlgray}{gray}{0.9}
\sethlcolor{vlgray}

\usetikzlibrary{shapes}

\RestyleAlgo{ruled}

\definecolor{SkyBlue}{RGB}{14, 118, 188}
\definecolor{BrightRed}{RGB}{223,82, 78}

\hypersetup{pdfborder = {0 0 0.5 [3 3]}, colorlinks = true, linkcolor = BrightRed, citecolor = SkyBlue}

\newcommand{\R}{\mathbb{R}}

\newcommand{\mycode}[1]{\texttt{#1}}
\newcommand{\berndist}[1]{\operatorname{Bernoulli}\left({#1}\right)}

\theoremstyle{plain}
\newtheorem{theorem}{Theorem}

\theoremstyle{definition}

\theoremstyle{Remark}
\newtheorem{remark}[theorem]{Remark}

\title{Pre-analysis protocol for an observational study on the effects of adolescent sports participation on health in early adulthood}
\author{Ajinkya H.~Kokandakar\thanks{University of Wisconsin--Madison}, Yuzhou Lin\thanks{Harvard University}, Steven Jin\thanks{University of Pennsylvania}, \\Jordan Weiss\thanks{Stanford University}, Amanda R.~Rabinowitz\thanks{Moss Rehabilitation and Research Institute}, Reuben A. Buford May\thanks{University of Illinois Urbana-Champaign}, \\ Dylan Small\footnotemark[3], and Sameer K.~Deshpande\footnotemark[1]}

\begin{document}
\maketitle

\begin{abstract}
We will study the impact of adolescent sports participation on early-adulthood health using longitudinal data from the National Study of Youth and Religion.
We focus on two primary outcomes measured at ages 23--28 --- self-rated health and total score on the PHQ9 Patient Depression Questionnaire --- and control for several potential confounders related to demographics and family socioeconomic status.
Comparing outcomes between sports participants and matched non-sports participants with similar confounders is straightforward.
Unfortunately, an analysis based on such a broad exposure cannot probe the possibility that participation in certain types of sports (e.g. collision sports like football or soccer) may have larger effects on health than others. 
    
In this study, we introduce a hierarchy of exposure definitions, ranging from broad (participation in any after-school organized activity) to narrow (e.g. participation in limited-contact sports).
We will perform separate matched observational studies, one for each definition, to estimate the health effects of several levels of sports participation.
In order to conduct these studies while maintaining a fixed family-wise error rate, we developed an ordered testing approach that exploits the logical relationships between exposure definitions.
Our study will also consider several secondary outcomes including body mass index, life satisfaction, and problematic drinking behavior. 
\end{abstract}

\section{Introduction}
\label{sec:introduction}
Studies have shown that sports participation has a positive impact on the physical, emotional, and mental well-being of youth \citep{Zarrett2018, Donaldson2006}. 
Many of these studies revolve around adolescents’ participation in school sponsored team sports programs and shed light on the influence of sports participation on students’ mental health \citep{Downward2011, Donaldson2006}. 
For instance, involvement in school sports during adolescence is a statistically significant predictor of lower depression symptoms, lower perceived stress, and higher self-rated mental health in young adults \citep{Jewett2014}. 
Moreover, greater participation in team sports prospectively predicts fewer symptoms of depression and anxiety at subsequent timepoints \citep{Graupensperger2021}, suggesting that sports participation has an impact beyond the time at which the young people compete. 
Further evidence of the benefits of sports participation are suggested by the fact that the subjective well-being level of the students who do not participate in school sports decreases as the grade of the student increases \citep{Malli2018}, and the subjective happiness levels of students differ depending on their active sports participation \citep{BingolBingol2020}. 
Hence, by and large, the literature reflects the positive value of sports participation on well-being.  

Still, there is evidence that the nature of sports participation (e.g.,\ individual versus team sports) along with the intensity (e.g.,\ competing in more than one sport or for significant amounts of time), and the type of sport has an impact on the well-being of adolescents. 
For instance, although individual sports athletes (e.g.,\ those competing in swimming, cross-country, fencing, track, etc.) suffer less from anxiety or depression than non-athletes, a greater proportion of individual sports athletes report anxiety or depression compared to team sports athletes \citep{Pluhar2019}. 
This may be due, in part, to the fact that team sports athletes have the social support of other team members who can ameliorate the kinds of performance pressures to which individual sport participants succumb \citep{Boone2006}. 
Like the impact that a particular sport can have on adolescents, the number of sports in which an adolescent competes can also have an impact on the benefits of sports participation. 
For instance, teens who participate in two or more sports benefit the most from their involvement with sports when compared to teens who only participate in one sport \citep{Zarrett2018}. 
Finally, the type of sport in which adolescents compete can have an impact. 
For instance, collision and contact sports athletes report fewer anxiety and depressive symptoms than do no- or limited-contact sports athletes \citep{Howell2020} but have more extensive orthopedic injury histories. 

Interestingly, many studies on adolescent sports participation focus on the contemporaneous benefits of sports.
That is, they study the extent to which adolescents are benefiting from sports, both on and off the field, during their playing careers \citep{Panza2020}. 
We know much less about the lasting positive benefits from youth sports participation.  
In the proposed study, we will evaluate well-being outcomes that extend beyond those associated with adolescent sports participation. 
Drawing on data from the National Study of Youth and Religion \citep[NSYR;][]{NSYR_design}, we explore later-life (specifically early adulthood) outcomes for youth that have participated in sports and evaluate the relative benefits and risks of participation in contact, limited-, and non-contact sports.  

The remainder of this protocol is organized as follows. 
We introduce the NSYR dataset and the baseline covariates in Section~\ref{sec:data_description}.
Then, in Section~\ref{sec:analysis_plan}, we outline the planned analysis, carefully describing the exposure definitions, matching methodology, and our tree-based testing-in-order procedure. 
We report the composition of our matched sets in Section~\ref{sec:matching_results} and perform a simulation study to determine the power of our planned analysis to detect different effect sizes in Section~\ref{sec:simulation}.
We conclude with a short discussion in Section~\ref{sec:discussion}.

\section{The NSYR dataset}
\label{sec:data_description}
We will use data collected as part of the National Study of Youth and Religion, a nationally-representative study examining religion and spirituality of American youth from adolescence into young adulthood.
Although the NSYR incorporated both telephone surveys and in-depth interviews of subjects and their parents, we will only analyze data collected from the telephone surveys.
These surveys were conducted in 2002--2003 (Wave 1) when the subjects were 13--17 years old; 2005 (Wave 2) when the subjects were aged 16--21; 2007--2008 (Wave 3) when the subjects were aged 18--24; and 2013 (Wave 4) when the subjects were aged 23--28.
Sports participation information was recorded in Wave 1.
See \citet{NSYR_design} for further details about the design and procedures for the NSYR.

\textbf{Primary and secondary outcomes.} 
We consider two co-primary outcomes, both measured in Wave 4: self-rated health and score on the PHQ-9 \citep{Kroenke2001}.
During Wave 4, NSYR participants were asked ``Overall, would you say your health is: (1) Excellent, (2) Very good, (3) Good, (4), Fair, or (5) Poor?''
We will dichotomize responses by combing ``excellent'', ``very good'', and ``good'' responses into one category (coded as 1) and combining ``fair'' and ``poor'' into another category (coded as 0).
Self-rated health, as captured by this survey question, is a widely used holistic indicator of health \citep{IdlerBenyamini1997}. 
Our dichotomization is quite common in and supported by previous research \citep{Manor2000}.

The PHQ-9 is a standard instrument to evaluate the severity of depressive symptoms that asks respondents how often they have been bothered by nine different problems like feeling tired, having little energy, or feeling down, depressed or hopeless.
Responses to each item are scored from zero (``not at all'') to three (``nearly every day'').
Item scores are then summed to produced a measure of depressive symptom severity that ranges from zero to 27, with lower scores corresponding to better mental health.
According to \citet{Kroenke2001}, scores of five or less on the PHQ-9 are typically associated with the absence of any depressive disorder while scores of 15 or higher usually indicate major depression. 
For our analysis, we use the total PHQ-9 score as our second co-primary outcome.

Our secondary outcomes include body mass index, self-rated physical well-being, an indicator of problematic drinking, and a composite life satisfaction scale score.
We will construct the latter two outcomes by aggregating responses to individual NSYR questions. 

\textbf{Inclusion \& exclusion criteria.} 
Wave 1 of the NSYR collected data from 3,370 respondents.
We will restrict our analyses to the 2,088 (62\%) of subjects with complete observations of the co-primary outcomes in Wave 4.

\textbf{Potential confounders.}
Since participation in sports and other activities is not randomly assigned, a simple comparison of average outcomes between groups of sports participants and non-participants is susceptible to confounding.
To overcome potential confounding, we adjust for several demographic, socioeconomic and educational characteristics of each subject in our study and their family measured at the baseline in Wave 1. 
Specifically, we adjust for subject age (at NSYR enrollment); gender; race; and the type of school the subject attended in Wave 1.
We additionally adjust for several factors related to each subject's family, including the census region in which they lived in Wave 1; family income; family structure; and parental education.
Table~\ref{tab:covariate_summary} summarizes the distributions of each potential confounder for all subjects in our study.
Appendix~\ref{sec:app_covariates} contains additional information about the construction of potential confounders, including the names of the specific NSYR variables and measurements used in our construction.

\begin{table}[!h]
\centering
\caption{Summary statistics for each of potential confounders included in our analysis. For continuous variables, we report means with standard deviations in parentheses. For categorical variables, we report percentages with totals in parentheses. }
\label{tab:covariate_summary}
\small
\begin{tabular}{lrlr}
    \hline
    Potential confounder & Mean   &   Potential confounder & Mean \\ \hline
    Age (years) &  15.1 (1.41) &     Family Income & \\
    Gender & &   \hspace{1em} 1\textsuperscript{st} quintile & 17\% (349)\\
        \hspace{1em} Male ($\%$)& 53\% (1106) &      \hspace{1em} 2\textsuperscript{nd} quintile & 24\% (512)\\
        \hspace{1em} Female ($\%$)& 47\%\phantom{(} (982) &    \hspace{1em} 3\textsuperscript{rd} quintile & 17\% (358)\\
    Race & &     \hspace{1em} 4\textsuperscript{th} quintile & 22\% (456)\\
        \hspace{1em} White ($\%$) & 73.4\% (1532)&    \hspace{1em} 5\textsuperscript{th} quintile & 20\% (413)\\
        \hspace{1em} Black ($\%$) & 11.7\%\phantom{1} (245)&   Family Structure & \\
        \hspace{1em} Hispanic ($\%$) & 8.8\%\phantom{1} (183) &     \hspace{1em} Two parent biological & 58\% (1208)\\
        \hspace{1em} Mixed ($\%$) & 1.6\%\phantom{11} (33) &     \hspace{1em} Two parent non-biological & 13\%\phantom{1} (267)\\
        \hspace{1em} Native American ($\%$) & 1.3\%\phantom{11} (27) &     \hspace{1em} Single Parent/Other  & 29\%\phantom{1} (613)\\
        \hspace{1em} Asian ($\%$) & 1.5\%\phantom{11} (32) &     Max. Parental Ed. & \\
        \hspace{1em} Islander ($\%$) & 0.3\%\phantom{111} (6) &     \hspace{1em} AA/vocational degree & 17.4\% (363)\\
        \hspace{1em} Other ($\%$) & 0.7\%\phantom{11} (14) &     \hspace{1em} BA/BS degree & 25.3\% (528)\\
        \hspace{1em} Missing ($\%$) & 0.8\%\phantom{11} (16) &     \hspace{1em} High school degree & 32.6\% (680)\\
    Census Region & &     \hspace{1em} Higher degree & 21.2\% (443)\\
        \hspace{1em} Northeast & 17\% (348) &     \hspace{1em} Other & 3.3\%\phantom{1} (68)\\
        \hspace{1em} Midwest & 25\% (518) &    \hspace{1em} Missing & 0.3\%\phantom{11} (6)\\
        \hspace{1em} South & 37\% (791) &    Type of School & \\
        \hspace{1em} West & 21\% (431) & \hspace{1em} Public school & 85.5\% (1786)\\
        & &      \hspace{1em} Private school & 10.4\%\phantom{1} (217)\\
        & &      \hspace{1em} Home school & 2.5\%\phantom{11} (53)\\
        & &      \hspace{1em} Other & 1.4\%\phantom{11} (29)\\
        & &      \hspace{1em} Missing & 0.1\%\phantom{111} (3)\\ \hline
\end{tabular}
\end{table}

\section{Analysis plan}
\label{sec:analysis_plan}
At a high level, we wish to understand the effect of sports participation in high school on later-life health using observational data.
A simple analysis proceeds by first classifying each subject into an exposed group (e.g.,\ sports participants or participants of a specific sport) or a control group (e.g.,\ subjects who played no sports or did not play a specific sport) and then comparing average outcomes between the two groups.
Because extracurricular activities are not randomly assigned, the exposed and control groups can systematically differ across multiple baseline sociodemographic, economic, and health-related variables that (a) are strongly predictive of participation and (b) affect later-life health. 
In other words, such simple comparisons may be confounded. 

\subsection{Defining the exposure} 
\label{sec:define_exposure}
While there are myriad ways to adjust for potential confounding like matching (see Section~\ref{sec:matching_inference}), there is a more fundamental issue: in the Wave 1 survey, the subjects only indicated the specific after-school activities (e.g., basketball or band) in which they participated.
What is the best way to define exposure and control? 

To illustrate, suppose we classify every subject who reported any sports participation into the exposed group and classify every subject who reported no sports participation to the control group.
Then the exposed group contains subjects who played a single non-collision sport like tennis as well as subjects who played multiple collision sports like football, wrestling, and soccer.
Similarly, the control group contains subjects who participated in non-sport extracurricular activities like band or theater, as well as subjects who participated in no activities at all.
Using such broad definitions of exposure and control ignores the heterogeneity contained within each group.
We are worried, in particular, about the possibility that (i) only some members of the exposed group experience much worse health outcomes compared to the controls than others but (ii) their large effects are ``washed out'' by the remaining exposed subjects who have similar outcomes compared to the controls.

If broad exposure and control definitions can ``mask'' effects experienced by only certain subsets of subjects, we might alternatively adopt narrower definitions.
For instance, we could define the exposure group to contain only those subjects who (i) reported playing a collision sport and (ii) did not report participation in any activity.
While narrow definitions avoid the problem of missing an effect experienced by only some subjects, they do not take full advantage of the available data.
In particular, adopting narrow exposure and control definitions necessarily limits the sample size of our analysis.
Simply put, broad definitions might yield higher power but yield smaller effect sizes while narrow definitions might yield larger effect sizes but lower power. 

\begin{figure}[h]
	\centering
	\includegraphics{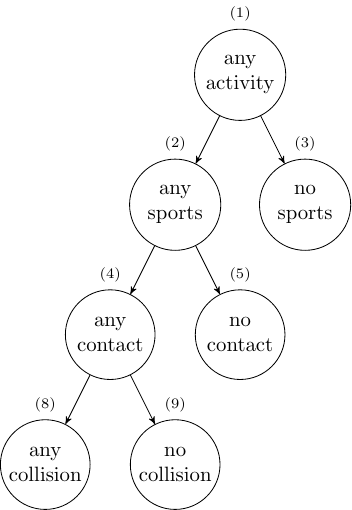}
\caption{Hierarchy of exposure definitions. We use the labels of the tree nodes (above, in parentheses) to index the hypotheses tested in Section~\ref{sec:ord_testing}.}
\label{fig:trt}
\end{figure}

In our study, there are no \textit{a priori} obvious choices of definitions for exposure and control.
Rather than picking one arbitrarily, we instead introduce multiple definitions that can be organized hierarchically using a binary tree (Figure~\ref{fig:trt}).

The broadest definition permitted by our dataset places all subjects who report participation in \textit{any} extracurricular activity (hereafter the ``any activity'' group) in the exposed group and all subjects who report no extracurricular activities (hereafter the ``no activity'' group) whatsoever in the control group.
As suggested above, although comparing the ``any activity'' to ``no activity'' groups potentially provides information about the benefits and risks of extracurricular activities, it provides little insight into the specific benefits and risks of sports participation.

Observe, however, that we can divide the ``any activity'' group into two non-overlapping subgroups, one containing those subjects who reported any sports participation (the ``any sports'' group) and those who only participated in non-sport activities (the ``no sports'' group).
Comparing both the ``any sports'' and ``no sports'' groups to the ``no activity'' group provides a more nuanced view into the benefits or harms conferred specifically by sports, relative to not participating in any after-school activity.
As indicated in Figure~\ref{fig:trt}, we can recursively subdivide the ``any sports'' group into ever-finer subsets.
We further subdivide the ``any sports'' group into ``any contact'' sports and ``non-contact'' sports groups.
This subdivision was motivated by concern that because higher rates of injuries among contact sports participants may mitigate any potential beneficial effects of sports participation.
We further subdivide the contact sports into collision sports (e.g., American football and soccer), vs non-collision contact sports (e.g., basketball).
This choice reflects our concern that certain sports involving high-impact collision, especially to the head and neck area, can lead to negative long-term health outcomes.

In forming the exposure groups in Figure~\ref{fig:trt}, we largely followed \citet{Meehan2016_division}'s definitions of collision and contact sports.
Briefly, collision sports are those in which body-to-body contact is legal and purposeful.
In contrast, contact sports as those in which body-to-body contact occurs, but purposeful body-to-body contact is not legal.
Finally, non-contact sports are those in which body-to-body contact is very rare.
Table~\ref{tab:sports_counts} lists all the collision, contact, and non-contact sports represented in our dataset, along with the total participants for each sport.
Because several NSYR subjects reported playing multiple sports, the sum of the counts in Table~\ref{tab:sports_counts} exceeds the total number of subjects in our study.
Although \citet{Meehan2016_division} does not consider diving, we classified diving as a collision sport.
This is because divers routinely make impact with the water at high-speeds and because they are at risk of collision with the diving platform, which is the leading cause of head and neck injury among adolescent divers \citep{Day2008_diving}.

\begin{table}[h]
    \centering
    \small
    \caption{Sports included in our dataset, with counts in parentheses.}
    \label{tab:sports_counts}
    \begin{tabular}{lll}
        \hline
        \textbf{Non-contact} &   \textbf{Contact} & \textbf{Collision} \\
        \hline
        Track (217) &   Basketball (509)   & Football (367) \\
        Volleyball (135) &   Soccer (263)  & Wrestling (80)\\
        Cross Country (70) &  Baseball (185) & Martial Arts (37)\\
        Tennis (64) &  Softball (120)    & Lacrosse (32)\\
        Swimming (59) &   Gymnastics (22) & Hockey (30)\\
        Golf (47) &   Field Hockey (16) & Boxing (6)\\
        Racquetball (3) &  Fencing (4)   & Diving (6)\\
        Crew (9) &  Flag Football (4)   & Rugby (1)\\
        &  Water Polo (3)  &  \\
        &   Roller Hockey (2)  & \\  
        \hline  
    \end{tabular}
\end{table}

\subsection{Testing in order}
\label{sec:ord_testing}

Having classified our study subjects into the ``no activity'' group or any of the groups shown in Figure~\ref{fig:trt}, it is tempting to carry out a matched observational study comparing each pair of non-overlapping groups.
For instance, we could compare the ``any contact'' and ``no contact'' groups to determine if only playing non-contact sports was safer than playing sports featuring any amount of contact.
Unfortunately, exhaustively comparing each pair of non-overlapping subgroups presents both statistical and practical challenges. 
Statistically, these comparisons introduce a multiple testing problem. 
Practically, our dataset is not large enough to guarantee that all such comparisons are powered to detect even large differences before correcting for multiple testing.

Which comparisons, then, ought we make? 
Can we select comparisons in a data-adaptive fashion while protecting against making too many false discoveries? 
As we describe below, it turns out that we can by using a tree-structured testing-in-order procedure.
Instead of comparing each pair of exclusive groups, we plan to compare each group to a common control group, the ``no activity'' group. 
In the context of Figure~\ref{fig:trt}, we will perform a matched observational study comparing the indicated exposure group to the ``no activity'' group at each node in the tree.
Rather than employing a simple Bonferroni correction to the seven hypothesis tests, we instead exploit certain logical relationships between the hypothesis.
Specifically, if we know that a hypothesis at one node is false, then it must be the case that at most one of the hypotheses at its children can be true. 

To illustrate this structure, let $\tau_{0} \in \R$ be a putative constant exposure effect and consider testing the following hypotheses
\begin{itemize}	
	\item $H_0^{(1)}: \tau^{(1)} = \tau_{0}$ (effect of participating in any activity vs no activity)
	\item $H_0^{(2)}: \tau^{(2)} = \tau_{0}$ (effect of participating in any sports activities vs no activity)
	\item $H_0^{(3)}: \tau^{(3)} = \tau_{0}$ (effect of participating in non-sports activities vs no activity)
	\item $H_0^{(4)}: \tau^{(4)} = \tau_{0}$ (effect of participating in contact sports'' vs no activity)
	\item $H_0^{(5)}: \tau^{(5)} = \tau_{0}$ (effect of participating in non-contact sports vs no activity)
	\item $H_0^{(8)}: \tau^{(8)} = \tau_{0}$ (effect of participating in collision sports vs no activity)
	\item $H_0^{(9)}: \tau^{(9)} = \tau_{0}$ (effect of participating in non-collision sports vs no activity)
\end{itemize}

We will, for the most part, be concerned with testing these hypotheses with $\tau_{0} = 0.$ 
That is, we first wish to identify which exposures, if any, have a non-zero effect relative to the common control condition of not participating in any after-school activities.
If a null hypothesis corresponding to particular node in the tree in Figure~\ref{fig:trt} is false, then at least one of the hypotheses corresponding to that nodes' children in the tree must also be false.
For instance, if there is truly a non-zero effect of participating in ``any activity'', then it cannot be the case that there is no effect of participating in either sports or non-sports activities. 
Formally, if $H_{0}^{(1)}$ with $\tau_{0} = 0$ is false, then at least of $H_{0}^{(2)}$ or $H_{0}^{(3)}$ with $\tau_{0} = 0$ is also false.
More abstractly, if a more coarsely defined exposure (e.g., ``any activity'') has a non-zero effect on the outcome, then \textit{at least} one of the exposures among a set of finer, mutually exclusive and exhaustive exposures (e.g., ``any sports'' and ``non-sports'') must have a non-zero effect on the outcome. 

To test multiple tree-structured hypotheses while controlling the probability of falsely rejecting even one true null hypothesis (i.e., controlling the family-wise error rate), we will use a testing-in-order procedure similar to the one described in \citet{Rosenbaum_testing}. 
Briefly, we always test $H_{0}^{(1)}$ and whenever we reject a hypothesis, we proceed to test the hypotheses at its children. 
If we fail to reject a hypothesis, however, we stop testing and do not test any of the descendant hypotheses.
For instance, if we reject $H^{(1)}_{0},$ then we are permitted to test both $H^{(2)}_{0}$ and $H^{(3)}_{0}$ with the same value $\tau_{0}.$
Further, if we subsequently reject $H^{(2)}_{0},$ we are permitted to test both $H^{(4)}_{0}$ and $H^{(5)}_{0}.$
But if we cannot reject $H_{0}^{(2)},$ the process stops and we do not test either $H_{0}^{(4)}$ or $H_{0}^{(5)}.$

\begin{figure}[!ht]
	\centering
	\begin{subfigure}{0.3\textwidth}
		\centering
		\includegraphics[page=1]{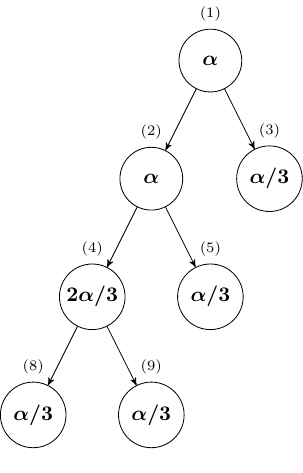}
        \caption{} 
        \label{fig:hyp1}
	\end{subfigure}
	\begin{subfigure}{0.5\textwidth}
		\centering
		\includegraphics[page=2]{figures/fig_alphas.pdf} 
        \caption{}
        \label{fig:hyp2}
	\end{subfigure}
	\caption{Ordered hypothesis testing strategy. (a) Significance level for each of the hypothesis tests. (b) Example scenario in which we reject $H^{(1)}_0$, $H^{(2)}_0,$ and $H_0:^{(5)}$ (gray nodes), fail to reject $H_{0}^{(3)}$ and $H_{0}^{(4)}$ (white nodes), and never test $H_{0}^{(8)}$ and $H_{0}^{(9)}$ (nodes with dashed outlines)}
    \label{fig:hyp}
\end{figure}

Figure~\ref{fig:hyp1} shows the levels at which we are able to perform the hypothesis tests while still maintaining a family-wise error rate of $\alpha$.
Basically, we must allocate the significance levels for each individual test in such a way that significance levels for all hypotheses that can simultaneously be true must add up to at most $\alpha$.
For instance, if $H^{(1)}_0$, $H^{(2)}_0$ and $H^{(4)}_0$ are all false, then at most one of $H^{(8)}_0$ and $H^{(9)}_0$ can be true. 
In either case, at most three hypotheses can be true (either $H^{(8)}_{0}$ or $H^{(9)}_{0}$ along with $H^{(3)}_0$ and $H^{(5)}_0$). 
Therefore, we can test all of them at a level of $\alpha/3$.
Each possible configuration of true hypotheses imposes a constraint on the level at which we are able to test each hypothesis. 
In Appendix~\ref{sec:app_testing_tree} we describe a strategy for setting our significance levels to satisfy these constraints.
This strategy is a specific instance of a more general strategy due to \citet{meijer2015dagtesting}.
It is also similar to a testing procedure introduced in \citet{meinhausen2008} but assigns different levels of significance to each test.

Figure~\ref{fig:hyp2} shows a potential realization of our testing strategy.
The gray nodes correspond to hypotheses that were tested and rejected, the white nodes corresponded to hypotheses that were tested but not rejected, and the nodes with dashed outlines correspond to untested hypotheses.
In the illustrated realization, because we rejected $H_{0}^{(1)},$ we proceeded to test both $H_{0}^{(2)}$ and $H_{0}^{(3)}.$
Upon rejecting $H_{0}^{(2)}$ we subsequently tested both $H_{0}^{(4)}$ and $H_{0}^{(5)}.$
However, because we failed to reject $H_{0}^{(4)},$ the procedure stops.

\subsection{Matching and randomization inference}
\label{sec:matching_inference}

We use matching followed by randomization inference to test each hypothesis considered in our testing-in-order procedure.

\textbf{Matching.}
We cannot directly compare the average outcomes in the exposed and control groups due to concerns about confounding. Matching accounts for confounding by only comparing the  average outcomes \textit{within} matched sets of individuals with similar values of the confounders.
In this study, we separately constructed matched sets for each exposure, with the set of units who participated in ``no activity'' as the common pool of controls.
We implemented optimal full matching  through the \textsf{R} package \textbf{optmatch} \citep{optmatch}  for each activity participation comparison to construct the matched sets. 
Every matched set had either one exposed subject and multiple control subjects, or multiple exposed subjects and one control subject. 
The full matching method imposes a propensity score caliper  and minimizes the rank-based Mahalanobis distance between matched subjects with similar propensity scores. 
In particular, we constructed the seven matches (one for each of the exposures in Figure~\ref{fig:trt}). 

To build the match at any one node in the tree, we first estimated covariate-balancing propensity scores \citep{ImaiRatkovic2014} using the \textbf{CBPS} \textsf{R} package \citep{cbps_package}.
Compared to other propensity score estimation procedures (e.g., gradient boosting or $L_{1}$-regularized logistic regression), we found that the CBPS estimates yielded better-balanced matched sets.
Before each match, we first dropped (i) the exposed subjects whose propensity scores are larger than the largest propensity scores among controls and (ii) the control subjects whose propensity scores are smaller than the smallest propensity score among the exposed. 
If a subject was dropped in one match, then they are also excluded from the subsequent matches. 
We report the number of subjects dropped prior to each match in Appendix~\ref{sec:app_treat}.

To check whether matching adequately balanced the exposed and control units in terms of the matched covariates within matched sets, we calculated the standardized differences before matching and after matching. 
The standardized difference of a covariate before matching is the difference in means of the covariate for the exposed vs. controls within group pooled standard deviation units. 
The standardized difference after matching is the weighted average of the difference in means within matched sets between the exposed and control units in the same within group pooled standard deviation units as before matching, where the weighting is by the number of exposed units in the matched set. 
To achieve adequate balance over the matched covariates, for every match we aimed to reduce the absolute value of the standardized difference for every matched covariate between the exposed and control units below 0.1 after matching.
More details about the matching algorithm and the computation of balance metrics are provided in Appendix~\ref{sec:app_matching_defs}.

For each exposure, we restricted the ratio of controls to exposed units between $1:k$ and $k:1$, with $k$ taking values $1, 2, \dots, 10.$
In this way, each matched set has a maximum size of $k+1$, with either (a) one exposed unit and a maximum of $k$ controls, or (b) one control and a maximum of $k$ exposed units. 
For the final matched sets, we chose the value of $k$ that minimized the number of covariates with absolute standardized difference in the interval $(0.1, 0.2)$ after matching. 
If there is any covariate for which the absolute standardized difference after matching exceeds $0.2,$ we declare the matching as failed.

\textbf{Randomization inference.}
Our goal is to estimate the effects of different levels of activity participation on our primary and secondary outcomes. 
We will conduct randomization inference on every outcome within each matched comparison.
For binary outcomes, we test the composite null of no effect described in \citet{fogarty_testing} (see Appendix~\ref{sec:app_binary_testing} for more details).
For continuous outcomes, we test the null hypothesis of no constant additive effect using the default  m-test implemented in the \texttt{senfm} function in the \textsf{R} package \textbf{sensitivityfull} \citep{Rosenbaum_sens}.
Note that although the PHQ9 score is integer-valued, we will treat it like a continuous outcome.
For each test, we will compute two-sided p-values when running our testing-in-order procedure.
For the sake of completeness, we will also report 95\% confidence intervals for every possible comparison, including those not reached in our testing-in-order procedure.
Code for forming the matched sets and running the testing-in-order procedure is available at \url{https://github.com/ajinkya-k/nsyr-matching-testing}.

\section{Matching results}
\label{sec:matching_results}
\begin{figure}[!h]
\centering
\includegraphics{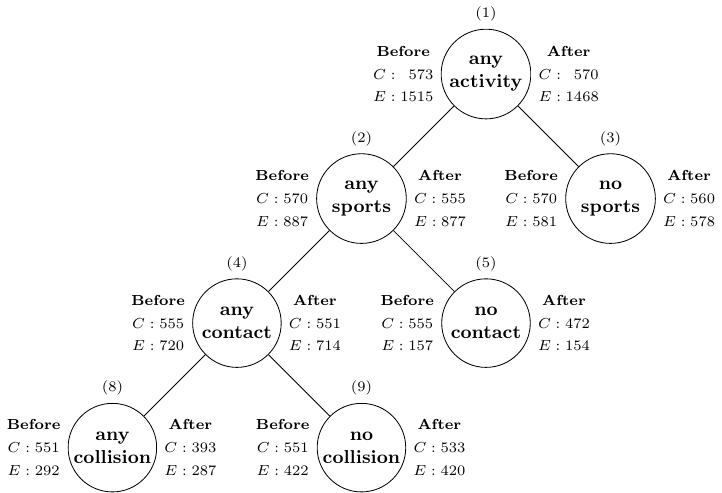}
\caption{Number of available exposed(E) and control (C) units before (left) and after (right) matching for each exposure}
\label{fig:matchtree}
\end{figure}

Figure~\ref{fig:matchtree} shows the sample size before and after matching for each of the exposures in the hierarchy. Note that the number of exposed units before matching for sibling exposures adds up to the number of matched exposed units for the parent exposures, and the number of controls available for matching is the same as the number of matched controls for the parent exposure. 

\textbf{Quality of Matches.}
In all but two cases, the standardized difference for the covariates after matching is within the range $[-0.1, 0.1]$.
In the case of the ``any activity'' exposure,  the standardized difference for the ``Northeast region'' binary indicator is 0.1032, which is very close to our threshold of $0.1$. 
In other words, subjects reporting participation in any activity were somewhat more likely to live in the Northeast census region than matched controls.
We do not feel that such imbalance will introduce substantial bias to our analysis. 

Secondly, the standardized difference for the ``Other race'' indicator was $0.17$ for the ``no contact'' exposure. Further examination revealed that there were only four units in this category available for matching.
Achieving better balance would require near exact balance on this covariate. 
When we attempted this, the quality of matches worsened for other covariates. 
Ultimately, we are not worried about the additional bias introduced by these four subjects to our analysis. 

\begin{remark}
Because we remove subjects with extreme propensity scores at each level, the causal conclusions we draw from each test generalize only to the overlap population at that level.
This is, in some sense, an inevitable limitation: including incomparable subjects in our analysis would lead to worse matches and potentially biased effect estimates.

\end{remark}

\section{Power simulation}
\label{sec:simulation}
It is not \textit{a priori} obvious that the ordered testing procedure described in Section~\ref{sec:ord_testing} is powered to detect small or large exposure effects.
To validate the testing procedure and to get a sense of its ability to detect true exposure effects, we performed an extensive semi-synthetic simulation study.
We specifically simulated outcomes for each subject in our dataset while keeping the covariates and matched sets fixed.
We perform two separate simulation studies, one with binary outcomes and one with integer outcomes.
In each study, we considered multiple data generating processes (DGP's) and created $2,500$ synthetic datasets for each DGP.
Throughout our simulation study, we tracked the following quantities for each hypothesis $H_{0}^{(t)}$ in the tree:
\begin{itemize}
\item{Marginal rejection rate: the proportion of simulations in which $H_{0}^{(t)}$ was rejected.}
\item{Testing rate: the proportion of simulations in which $H_{0}^{(t)}$ was tested. }
\end{itemize}
Note that in order to reject a hypothesis, we must first test it and obtain a p-value less than the associated significance level shown in Figure~\ref{fig:hyp1}.
Since $H_{0}^{(t)}$ is tested if and only if its parent hypothesis is rejected, the testing rate for $H_{0}^{(t)}$ is exactly the same as the marginal rejection rate of its parent hypothesis.

For each DGP, to generate outcomes, we specified a constant treatment effect within each of five mutually exclusive subsets of our analysis sample:
\begin{itemize}
\item{$\mathcal{S}^{(0)}$: control subjects, who do not participate in any activity at all}
\item{$\mathcal{S}^{(3)}$: ``no sports'' group}
\item{$\mathcal{S}^{(5)}$: ``no contact'' sports}
\item{$\mathcal{S}^{(8)}$: ``any collision'' sports}
\item{$\mathcal{S}^{(9)}$: ``no collision'' sports}
\end{itemize}


For ease of exposition, we first discuss simulation for integer outcomes that mimic the PHQ-9 score followed by simulation for binary outcomes.
For each of the two outcomes we present the results for three simulation scenarios: 
\begin{itemize}
	\item Adverse effects of playing collision sports 
	\item Beneficial effects of playing any sports
	\item Beneficial effects of non-contact sports, adverse effects of collision sports, and no effect of contact sports with no collisions 
\end{itemize}

\subsection{Simulation with integer outcomes}

To mimic the PHQ-9 score, we simulate potential outcomes under exposure and control that are an integer between 0 and 27.
To simulate the potential outcome under control we first generate a latent potential outcome $\tilde{Y}_j(0)$ from a normal distribution with mean 13 and standard deviation of 9 and then forming the actual potential outcome $Y_{j}(0)$ by rounding $\tilde{Y}_{j}(0)$ to the closest integer between 0 and 27. 
Then, we set the potential outcome under exposure $Y_{j}(1)$ equal to the closest integer in $\{0, 1, \ldots, 27\}$ to the value $Y_{j}(0) + \mathbbm{1}\left(j \in \mathcal{S}^{(t)}\right) \times \tau^{(t)}.$
We introduce truncation when simulating both potential outcomes to mimic the fact that PHQ-9 scores are integer-valued. 
Consequently, in our simulated data, the actual average of $Y_{j}(1) - Y_{j}(0)$ within group $\mathcal{S}^{(t)}$ sometimes deviated slightly from the nominal treatment effect $\tau^{(t)}.$
However, these deviations tended to be small: for small and medium $\tau,$ the deviations tended to be less than half a point and for larger $\tau,$ the deviations tended to be less than three-quarters of a point. 
Finally, the simulated observed outcome for each unit $Y_j$ is just the potential outcome under the observed exposure.
For each simulated dataset, we ran our tree based testing procedure with $\tau_{0} = 0$ to determine which exposure had non-zero effects.

\textbf{Simulation 1: Adverse effect of collision sports on mental health}

To simulate an adverse effect for playing collision sports, we first set $\tau^{(3)} = \tau^{(5)} = \tau^{(9)} = 0$, while we simulate data for the ``any collision'' exposure with $\tau^{(8)} \in \{1, 3, 5\}.$
Note that larger PHQ-9 scores correspond to worse mental health. 
In this simulation, we had $Y_j(1) = Y_j(0)$ for all subjects who did not participate in any collision sport (i.e. $j \notin \mathcal{S}^{(8)}$).

\begin{figure}[!ht]
	\centering
	\captionsetup[subfigure]{justification=centering}
	\begin{subfigure}[b]{0.49\textwidth}
		\includegraphics[page=1,width=\textwidth]{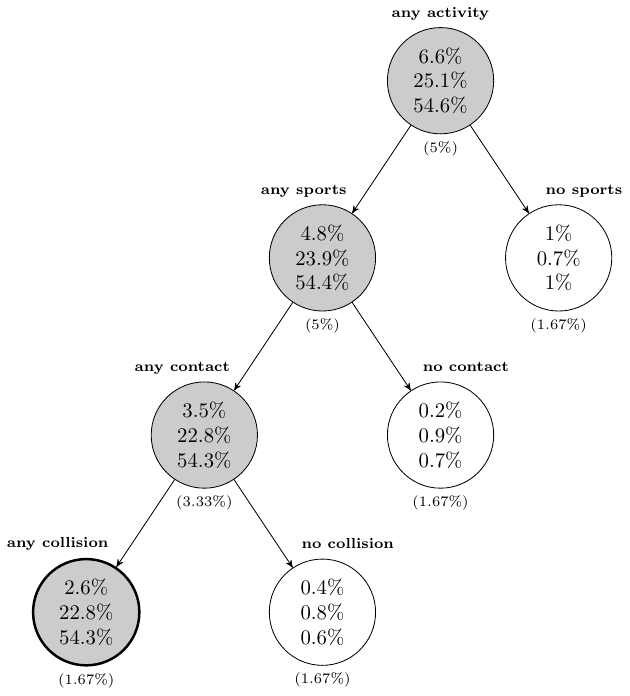}
		\centering
		\textbf{FWER:} $1.76\%$, $2.28\%$, $2.68\%$
		\subcaption{}
		\label{fig:cont_sim_ac}
	\end{subfigure}
	\hspace*{\fill}
	\begin{subfigure}[b]{0.49\textwidth}
		\includegraphics[page=2,width=\textwidth]{figures/fig_power_cont1.pdf}
		\centering
		\textbf{FWER:} $0.96\%$, $1.2\%$, $1.68\%$
		\subcaption{}
		\label{fig:cont_sim_as}
	\end{subfigure}
	\caption{Empirical rejection rates for (a) Simulation 1, adverse effect of ``any collision'' ($\tau^{(8)} \in \{1, 3, 5\}$); (b) Simulation 2, beneficial effect of ``any sports'' ($\tau^{(2)} \in \{-1, -3, -5\}$). The values in the nodes denote the marginal rejection rates for the three effect sizes respectively. The gray color denotes the nodes where the null hypothesis of no effect is false, and thick border denotes the nodes where the effect was applied.}
	\label{fig:cont_sim}
\end{figure}

Figure~\ref{fig:cont_sim_ac} shows the results of 2500 simulations for three exposure effect sizes for  ``any collision'' -- small effect ($\tau^{(8)} = 1$), medium effect ($\tau^{(8)} = 3$), and large effect ($\tau^{(8)} = 5$).
The values inside the node are the marginal rejection rates.
The family wise error rate (FWER) reported is computed as the proportion of simulations in which any of the three true null hypotheses --- ``no sports'', ``no contact'' or ``no collision'' --- was rejected.
Interestingly, the FWER across all three effect sizes is considerably less than the targeted 5\% level, suggesting that our testing-in-order procedure can be rather conservative in the presence of non-zero effects.
In addition to this, the rejection rates for each of the true null hypotheses is less than the corresponding significance levels assigned to the tests. 

It is clear from Figure~\ref{fig:cont_sim} that the power to detect the effect of ``any collision'' increases with the size of the effect, $\left \lvert\tau^{(8)}\right\rvert$.
We additionally observe increased power for detecting the effect of ``any contact'' sports, ``any sports'' and ``any activity'' as the exposure effect of ``any collision'' increases.
The marginal power for ``any collision'' is low when $\left\lvert\tau^{(8)}\right\rvert = 1$, because our procedure reaches the test for ``any collision'' only 4\% of the time. 
The low testing rates can be explained by the fact that the effect only manifests in the small subset of units that engaged in ``any collision'' sports, which is mostly washed out by the zero effect of other exposures.
Essentially, we did not reject the null hypotheses for ``any activity'', ``any sports'' and ``any contact'' often enough to be able to test for an effect of ``any collision''.


\textbf{Simulation 2: Beneficial effects of sports on mental health}

To simulate beneficial effects for any sports, we first set the no sports exposure effect, $\tau^{(3)} = 0$, while we simulate data for the all the sports exposures with $\tau^{(5)} = \tau^{(8)} = \tau^{(9)} \in \{-1, -3, -5\}$, because lower PHQ-9 scores denote better mental health.
Since we use the same effect for all leaf node exposures under any sports it is equivalent to write: $\tau^{(2)} \in \{-1, -3, -5\}$.

Figure~\ref{fig:cont_sim_as} shows the results for 2500 simulations using this setup.
As in the first scenario, the FWER is well below the $5\%$ threshold for all three effect sizes.
The rejection rate for the true null hypothesis (no effect of ``no sports'') coincides with the FWER as there is only one true null hypothesis, and is less than the corresponding significance level of $1.67\%$.
As the exposure effect of any sports $\tau^{(2)}$ increases, the marginal rejection rate for any sports and all its descendant exposures also increases.
Notice that rejection rates when $|\tau^{(2)}| = 1$ is almost 4 times that of the corresponding rejection rates in Simulation 1. 
This is because the ``any sports'' cohort is much larger than the ``any collision'' cohort. 
Therefore, even small effect sizes are enough to detect the effect of ``any sports''.
Observe that we reject the false null hypotheses almost every time when  $\tau^{(2)} = -5$.

\textbf{Simulation 3: Beneficial effects of non-contact sports and harmful effects of contact sports on mental health}

In this scenario, we first set the exposure effect of ``any collision'' with $\tau^{(8)} \in \{1, 3, 5\}$, and we choose two different sets of effect sizes for ``no contact'' sports, $\tau^{(5)} = - \tau^{(8)}$ and $\tau^{(5)} = -0.5 \tau^{(8)}$, while all other exposures have zero effect.

\begin{figure}[!h]
	\centering
	\captionsetup[subfigure]{justification=centering}
	\begin{subfigure}[b]{0.49\textwidth}
		\includegraphics[page=1, width=\textwidth]{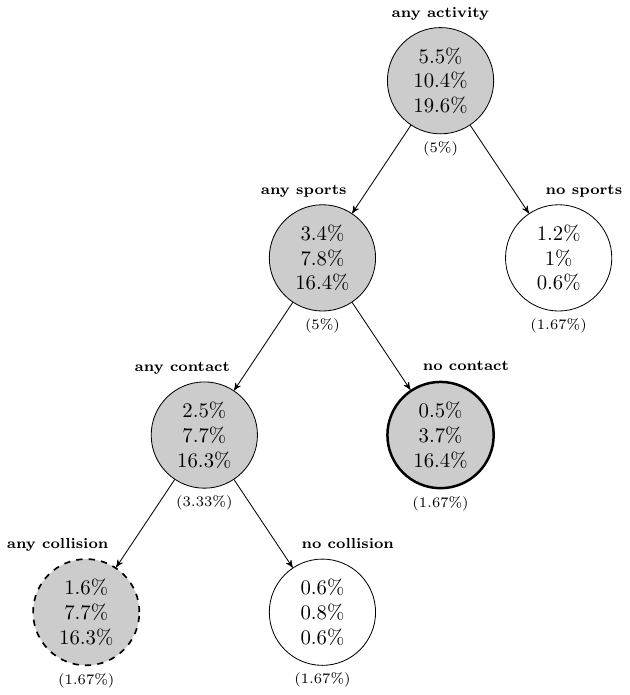}
		\centering
		\textbf{FWER:} $1.68\%$, $1.72\%$, $1.12\%$
		\subcaption{}
		\label{fig:cont_mix_a}
	\end{subfigure}
	\hspace*{\fill}
	\begin{subfigure}[b]{0.49\textwidth}
		\includegraphics[page=2, width=\textwidth]{figures/fig_power_cont2.pdf}
		\centering
		\textbf{FWER:} $1.08\%$, $1.76\%$, $1.72\%$
		\subcaption{}
		\label{fig:cont_mix_b}
	\end{subfigure}
	\caption{Empirical rejection rates when the exposure effect for sports are mixed. The values in the nodes denote the marginal rejection rate (i.e. fraction of runs in which the null hypothesis at the node was rejected). The gray color denotes the nodes where the null hypothesis of no effect is false.}
	\label{fig:cont_sim_mix}
\end{figure}

The results for the two sets of simulations are shown in Figure~\ref{fig:cont_mix_a} for $\tau^{(5)} = - \tau^{(8)}$ and in Figure~\ref{fig:cont_mix_b} for $\tau^{(5)} = - 0.5\tau^{(8)}$.
The FWER is computed as the proportion of simulations in which the at least one of two true null hypotheses of no-effect for ``no sports'' and ``no collision'' are rejected.
As before we observe that the FWER for all simulations is lower than 5\%.
Secondly, the rejection rates for the true null hypotheses of no effect for ``no collision'' sports and ``no sports'' activities are less than the corresponding significance levels.

The marginal power for both exposures is roughly similar when both the effect sizes are small ($\tau^{(8)}$ = 1).
For the same size of the ``any collision'' exposure effect, we can detect this effect more often if the ``no contact'' exposure effect has a smaller magnitude.
This is to be expected, because when these exposures have effects of the same magnitude but opposite signs, they negate each other, and it becomes harder to reject the no-effect null hypotheses for ``any activity'' and ``any sports'', which are the common ancestor exposures. 
In fact, when $\tau^{(8)} = -2\tau^{(5)}$, the rejection rates for all the ancestor nodes of ``any collision'' have about twice the rejection rate when compared to those when $\tau^{(8)} = -\tau^{(5)}$.
Despite the rejection rate for ``any sports'' being high, the rejection rate for ``no contact'' is low when $\tau^{(5)}$ has a lower magnitude than $\tau^{(8)}$.
This seems to point to the possibility that having exposure effects of differing magnitudes but opposite signs entails lower power for the exposure with lower magnitude effects, but much increased power for those with the higher magnitude.

\subsection{Simulation with binary outcomes}
Next, we consider the simulations with binary outcomes.
For each unit $j$ in the dataset (including all exposure and control groups), we generate the potential outcome under control as:
$
	Y_j(0) \sim \berndist{\pi^{(0)}}.
$
For the control group, we set $Y_j(1) = Y_j(0)$.
Next, within each exposure group $t \in \{2, 5, 8, 9\}$, we generate the potential outcome under exposure as: 
$
    Y_{j}(1) \sim \berndist{\pi^{(t)}} \, \text{ for } j \in \mathcal{S}^{(t)},
$
where $\pi^{(t)}$ is the probability of self-reporting good health.
Finally, the observed outcome $Y_j$ is the same as the potential outcome under the observed exposure (or control).

Next, we use our ordered testing to choose and test the hypotheses in an adaptive fashion. 
For any exposure $t$ (control excluded) chosen by the procedure, we consider the hypotheses: 
\begin{align*}
    H_0: \pi^{(t)} - \pi^{(0)} = 0\quad \text{vs} \quad
    H_1: \pi^{(t)} - \pi^{(0)} \neq 0\ , 
\end{align*}
and we use the test developed in \cite{fogarty_testing}.

\textbf{Simulation 4: Adverse effect of collision sports on self-reported health}

To simulate adverse effects for playing collision sports, we first set $\pi^{(3)} = \pi^{(5)} = \pi^{(9)} = 0.5$, that is all exposures groups other than ``any collision'' have no effect.
On the other hand, we use $\pi^{(8)} \in \{0.1, 0.3, 0.4\}$, to simulate data with adverse effects for the ``any collision'' exposure. 

\begin{figure}[!h]
	\centering
	\captionsetup[subfigure]{justification=centering}
	\begin{subfigure}[b]{0.49\textwidth}
		\includegraphics[width=\textwidth,page=1]{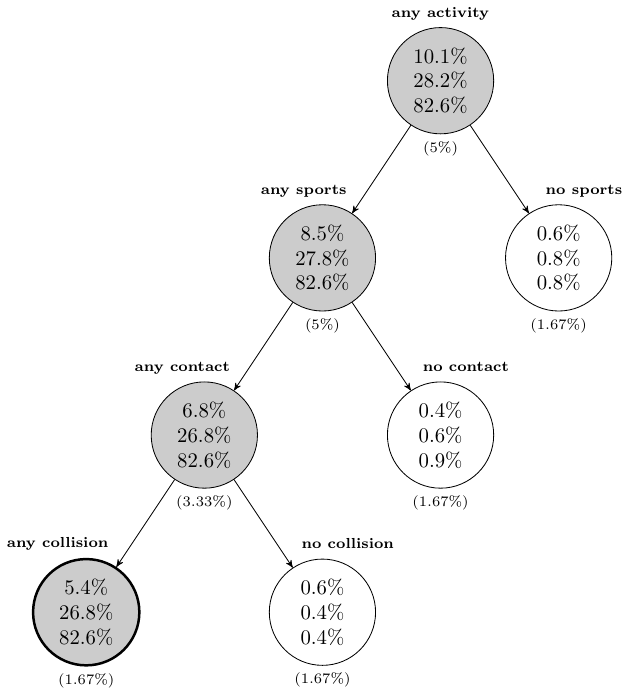}
		\centering
		\textbf{FWER:} $1.36\%, 1.84\%, 2.04\%$
		\subcaption{}
		\label{fig:bin_sim_ac}
	\end{subfigure}
	\hspace*{\fill}
	\begin{subfigure}[b]{0.49\textwidth}
		\includegraphics[width=\textwidth,page=2]{figures/fig_power_bin1.pdf}
		\centering
		\textbf{FWER:} $0.44\%, 0.64\%, 1.32\%$
		\subcaption{}
		\label{fig:bin_sim_as}
	\end{subfigure}
	\caption{Empirical rejection rates for (a) Simulation 4, adverse effect of ``any collision'' ($\pi^{(8)} - \pi^{(0)} \in \{-0.1, -0.2, -0.4\}$); (b) Simulation 5, beneficial effect of ``any sports'' ($\pi^{(2)} - \pi^{(0)} \in \{0.1, 0.2, 0.4\}$). The values in the nodes denote the marginal rejection rates for the three effect sizes respectively. The gray color denotes the nodes where the null hypothesis of no effect is false, and thick border denotes the nodes where the effect was applied.}
	\label{fig:bin_sim}
\end{figure}

Figure~\ref{fig:bin_sim_ac} shows the results of 2500 simulations for three exposure effect sizes for  ``any collision'' -- small effect ($\pi^{(8)} = 0.4$), medium effect ($\pi^{(8)} = 0.3$), and large effect ($\pi^{(8)} = 0.1$)
We simulated 2500 datasets for each effect size.
The FWER is computed as the proportion of simulations in which at least one of the true null hypotheses for ``no sports'', ``no contact'' and ``no collision'' exposures was rejected.
Similar to the simulation for integer outcomes, the FWER is much lower than the 5\% significance level.
The marginal rejection rates for the true null hypotheses are less than the corresponding significance levels.
The values inside the node are the marginal rejection rates i.e. the proportion of simulations in which we reject the null hypothesis.
It is clear from the figure that the ``power'' to detect the effect of ``any collision'' increases with the size of the effect, $\left|\pi^{(8)} - \pi^{(0)}\right|$.
In general, the marginal rejection rates are higher in this case as compared to the rejection rates in the corresponding integer outcome simulations.

\textbf{Simulation 5: Beneficial effects of sports on self-reported health}

To simulate a beneficial effect of participating in ``any sports'', we set the probability of self-reported good health for non-sports activity and the control as follows: $\pi^{(3)} = \pi^{(0)} = 0.5$.
We set $\pi^{(t)} \in \{0.6, 0.7, 0.9\}$ for all sports exposures $t \in \{5, 8, 9\}$.
Equivalently, we can say that $\pi^{(2)} \in \{0.6, 0.7, 0.9\}$ as the exposure effect is the same for all the leaf nodes under the any sports exposure.

Figure~\ref{fig:bin_sim_as} shows the results for the simulation.
In this case both the FWER and the marginal rejection rate for the true null hypothesis for ``no sports'' coincide, and the rejection rate is lower than the allocated significance level.
As before, we observe that the marginal rejection rate for the false null hypotheses increases as the exposure effect $\pi^{(t)} - \pi^{(0)}$ increases.
The marginal rejection rate for the no sports exposure is less than the $\alpha_3 = 1.67\%$.

\textbf{Simulation 6: Beneficial effects of non-contact sports, but non-beneficial effects of contact sports on self-reported health}

\begin{figure}[!ht]
	\centering
	\captionsetup[subfigure]{justification=centering}
	\begin{subfigure}[b]{0.49\textwidth}
		\includegraphics[page=1, width=\textwidth]{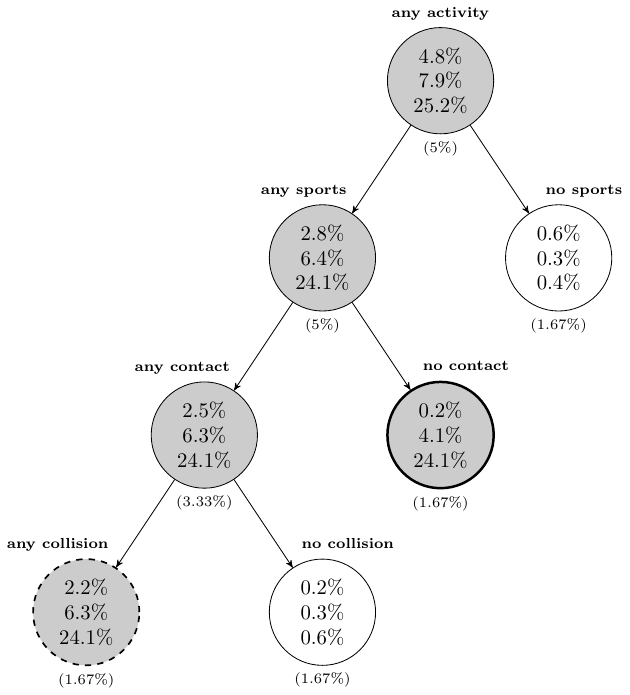}
		\centering
		\textbf{FWER:} $0.80\%, 0.60\%, 0.88\%$
		\subcaption{}
		\label{fig:bin_mix_a}
	\end{subfigure}
	\hspace*{\fill}
	\begin{subfigure}[b]{0.49\textwidth}
		\includegraphics[page=2, width=\textwidth]{figures/fig_power_bin2.pdf}
		\centering
		\textbf{FWER:} $1.08\%, 0.96\%, 1.08\%$
		\subcaption{}
		\label{fig:bin_mix_b}
	\end{subfigure}
	\caption{Empirical rejection rates when the exposure effect for sports are mixed, and the outcome is binary. The values in the nodes denote the marginal rejection rate (i.e. fraction of runs in which the null hypothesis at the node was rejected). The gray color denotes the nodes where the null hypothesis of no effect is false.}
	\label{fig:bin_sim_mix}
\end{figure}

In this case, we first simulate a non-beneficial effect of ``any collision'' by choosing $\pi^{(8)} - \pi^{(0)} \in \{-0.1, -0.2, -0.4\}$.
Then we simulate a beneficial effect of ``no contact'' sports with two different sets of effect sizes: $\pi^{(5)} - \pi^{(0)} = - (\pi^{(8)} - \pi^{(0)})$ and $\pi^{(5)} - \pi^{(0)} = - (\pi^{(8)} - \pi^{(0)}) / 2$, while all other exposures have zero effect.

Figure~\ref{fig:bin_sim_mix} shows the results of the simulations for the two effect sizes.
The FWER is computed as the proportion of simulations in which at least one of the two true null hypotheses of no-effect corresponding to ``no sports'' and ``no collision'' is rejected.
The FWER is much lower that the 5\% significance level, and the rejection rate for both the true null hypotheses is lower than the corresponding significance levels
Secondly, the marginal rejection for all null hypotheses increases as the effect sizes increase.
Notably, the rejection rate for all false null hypotheses is higher when the size of the effects for the two exposures of ``any collision'' and ``no contact'' are different i.e. for the scenario in which $\pi^{(5)} - \pi^{(0)} = - (\pi^{(8)} - \pi^{(0)}) / 2$.
This seems to point to the possibility that the exposure effects are hard to detect when they have opposite signs but the same magnitude.

\section{Discussion}
\label{sec:discussion}
We proposed an observational study of the effects of sports participation in adolescence on several health outcomes in early adulthood.
Rather than commit to one broad or one narrow definition of exposure, we introduced a hierarchy of exposures, which could be organized along a binary tree.
We will use a tree-based testing-in-order procedure that enables us to perform a matched observational studies for each exposure definition while maintaining a fixed family-wise error rate.
Using a comprehensive simulation study, we found that our procedure is able to detect small effects of broad exposures (e.g., playing any sports) as well as large effects of narrow exposures (e.g., playing any collision sport).

\clearpage
\bibliography{sports_participation_refs}

\clearpage
\appendix
{\Huge \textbf{Appendix}}

\appendix
\section{Baseline covariates}
\label{sec:app_covariates}
In this appendix, we detail the variable and how they are defined using the original survey. 

\begin{enumerate}
	\item 

\textbf{Age of the subject:} Measured in years, and imputed if missing.

\item \textbf{Family income:}
Family income quantile. 
Derived from \mycode{725) PINCOME} in the Wave 1 codebook. In the survey, household income was recorded based on intervals as follows: 
5000 (for income between \$0-\$10,000), 15000 (for income between \$10,000 and \$20,000), \dots, 95000 (for income between \$90,000 and \$100,000) and 105,000 (above \$100,000). We divided our subjects into quintiles based on the NSYR-recorded income levels. 

\item \textbf{Family structure:}
Describes the living situation for the subject, and can be one of three values: ``Two Parent Nonbiological'', ``Two Parent biological'', or ``Single Parent/Other''.
This is computed from variables \mycode{11)} - \mycode{16)} in the Wave 1 codebook. 

\item \textbf{Highest education of parents in household:} 
Derived from variables \texttt{735)} - \texttt{736)} in the Wave 1 codebook.
Maximum education level of household resident who assumes a mother role (biological, adoptive, stepparent) and household resident who assumes a father role (biological, adoptive, stepparent).  If the child is living with a single parent, then this is just the education level of that single parent. The possible values for this factor are: AA/vocational degree, BA/BS degree, High school degree, Less than high school, and missing.

\item \textbf{Gender of the subject:}
Recorded as the response to the question: ``Is (your teen) a boy or a girl?''. 
If the respondent refused the survey assumed the child is a boy.
For source see \mycode{10) TEENSEX} and \mycode{916) I-GENDER} in the Wave 1 codebook. The values are coded as $0$ for male, and $1$ for female.

\item \textbf{Race of the subject:} 
For source see \mycode{726) TEENRACE} in the Wave 1 codebook. The possible values are White, Black, Hispanic, Asian, Islander, Native American, Mixed, other, and missing. 
The missing category includes missing values and also respondents who do not know or refused to answer the question.

\item \textbf{Type of school the subject is attending:}
For source, see \mycode{101) PSCHTYP} in the Wave 1 codebook.
The type of school can be public, private, homeschooled, other, or missing (which means the respondent either did not know or refused to answer the question).
Note that the ``other'' category mergers five different responses into one category. 

\item \textbf{Census Region:}
For the source, see \mycode{901) CENREG} in the Wave 1 codebook. The values are coded as $1$ for Northeast, $2$ for Midwest, $3$ for South, and $4$ for West.
\end{enumerate}

\section{Exposure details}
\label{sec:app_treat}

Table~\ref{tab:composition} lists the number of exposed units, control units and the sample size for each of the exposures. Notice that the number of controls is the same, as we use ``no activity'' i.e. \mycode{any\_activity = 0} as the common pool of controls.
\begin{table}[!h]
	\centering
	\caption{Composition of samples for each of exposures}
	\label{tab:composition}
	\begin{tabular}{rccc}
		\hline
		\textbf{Exposure} &  \textbf{exposed} & \textbf{controls} & \textbf{total} \\
		\hline
		any activity & 1515 & 573 & 2088\\
		any sports & 921 & 573 & 1494\\
		no sports & 594 & 573 & 1167\\
		any contact & 757 & 573 & 1330\\
		no contact & 164 & 573 & 737\\
		any collision & 303 & 573 & 876\\
		no collision & 454 & 573 & 1027\\
		\hline
	\end{tabular}

\end{table}

When matching, we eliminated exposed and control units that had extreme propensity scores. A control is dropped if its propensity score was lower than the exposed unit with the least propensity score. Similarly, an exposed unit was dropped if its propensity score was higher than the control unit with the highest propensity score. See Table \ref{tab:matched_sample} for details. 

\begin{table}[!h]
	\centering
	\footnotesize
	\caption{Sample Sizes after matching}
	\label{tab:matched_sample}
	\begin{tabular}{rccccccccc}
		\hline
		 & \multicolumn{3}{c}{\textbf{Before Matching}} & \multicolumn{3}{c}{\textbf{Extreme propensity score}} & \multicolumn{3}{c}{\textbf{After matching}}\\
		\hline
		Exposure &  exposed & controls & total & exposed & controls & total & exposed & controls & total\\
		\hline
		any activity & 1515 & 573 & 2088 & 47 & 3 & 50 & 1468 & 570  & 2038\\
		any sports & 921 & 573 & 1494 & 10 & 15 & 25 & 877 & 555  & 1432\\
		no sports & 594 & 573 & 1167 & 3 & 10 & 13 & 578 & 560  & 1138\\
		any contact & 757 & 573 & 1330 & 6 & 4 & 10 & 714 & 551  & 1265\\
		no contact & 164 & 573 & 737 & 3 & 83 & 86 & 154 & 472  &  626\\
		any collision & 303 & 573 & 876 & 5 & 158 & 163 & 287 & 393  &  680\\
		no collision & 454 & 573 & 1027 & 2 & 18 & 20 & 420 & 533  &  953\\
		\hline
	\end{tabular}

\end{table}

\section{Matching methodology and balance metrics}
\label{sec:app_matching_defs}
\subsection{Matching algorithm details}
For each exposure, we use full matching on  the set of available exposed and control units. The control units included in the matched set for the parent exposure is the available set of controls. The available exposed units are the matched exposed units for the parent exposure that have the current exposure status as well-defined.

Given an exposure, we use a Mahalanobis distance with a propensity score caliper of $0.2$ times the standard deviation and with a caliper penalty of $1000$. For $k \in \{1,2, \dots, 10 \}$ we compute the full match by restricting the exposed to controls ratio between $1:k$ and $k:1$, and choose the value of $k$ that minimizes the number of covariates with absolute standardized difference between $0.1$ and $0.2$. We declare the matching as failed if for all $k$ there is some covariate with absolute standardized difference larger than or equal to $0.2$. 

\subsection{Balance metrics}
Let $x_{ij}$ denote the $i^{th}$ baseline covariate value for the $j^{th}$ unit, then the standardized difference for the $i^{th}$ covariate is: 
$$
	\delta_i = \frac{\sum_{j \in \text{exposed}} w_j \cdot x_{ij} - \sum_{j \in \text{controls}} w_j \cdot x_{ij}}{s_{i, pool}}
$$
with $s_{i, pool} = \sqrt{(s_{i,1}^2 + s_{i,0}^2)/2}$, where $s_{i,1}^2$ and $s_{i,0}^2$ are the variances of $i^{th}$ covariate for the exposed and control sub-samples before matching respectively.

To compute $\delta_i$ before matching we use $w_j = 1/n$ for all units ($n$ is the total number of units). The weights after matching are computed from the matched sets. 
For a matched set indexed by $k$ let $n_{k}^{(0)}$ and $n_{k}^{(1)}$ denote the number of control and exposed units in the matched set, then the (non-normalized) weight for each control unit is:
$
	v_j = n_{k}^{(1)}/n_{k}^{(0)}
$
for each control $j$ in matched set $k$.
Finally, we normalize the weights across the groups so that $w_j = 1/n^{(0)}$ for the matched controls and $w_j = v_j/\sum_{j \in \text{matched exposed}} v_j$ for the matched exposed units.
Finally, the absolute standardized difference for the $i^{th}$ covariate is simply $|\delta_i|$.

\section{Randomization inference for binary outcomes}
\label{sec:app_binary_testing}
Randomization inference usually assumes a constant additive exposure effect, but this is not appropriate for binary outcomes, because the exposure response tends to be heterogeneous \citep{fogarty_testing} when the outcome is binary.
This is because if the outcome under control itself is $1$, the constant additive effect of $1$ cannot be true for this unit, and similarly if the outcome under control is $0$, then the constant additive effect of $-1$ cannot be true.
Moreover, the assumption of additive effect (say $\tau$) defined in terms of proportion of subjects with good health i.e. $\pi^{(1)} - \pi^{(0)} = \tau$  where $\pi^{(0)}$ and $\pi^{(1)}$ are the true proportion of outcome being 1 under control and under exposure (in this case any activity) respectively, is also usually untenable.
This is because for any value of the additive effect (between 0 and 1), we can find some value of $\pi^{(0)}$ such that $\pi^{(0)} + \tau > 1$ (or $\pi^{(0)} + \tau < 0$ if $\tau$ is negative).

More concretely, we test the difference in proportion of good outcomes in the exposure as compared to the control group for each of the exposures $t$.
That is we test:
\begin{align*}
    H_0&: \pi^{(t)} - \pi^{(0)} = 0\\
    H_1&: \pi^{(t)} - \pi^{(0)} \neq 0\ .
\end{align*}
When performing randomization inference for binary data, there are multiple potential outcome schedules that correspond to the null of no effects, which can be expressed as the same probability of the outcome taking value 1 in both the exposed and the control groups, therefore \cite{fogarty_testing} suggest using a composite null.
We use the hypothesis test for the difference in proportion (termed ``risk difference'' in their paper) laid out in \cite{fogarty_testing}.
To combat the issue of multiple compatible potential outcome schedules, the p-value computed by the test corresponds to the worst case p-value among all possible potential outcome consistent schedules consistent with the null hypothesis, which is computed using an integer linear program.

\section{Tree based testing}
\label{sec:app_testing_tree}
The key challenge in developing this tree-structured hypotheses testing procedure is the allocation of the significance level for each of the tests. 
Note that we test $H_0^{(2)}$ and $H_0^{(3)}$ \textit{only if} $H_0^{(1)}$ has already been rejected. 
Now suppose that all the null hypotheses are in fact true, but we reject $H_0^{(1)}$ and $H_0^{(3)}$ but fail to reject $H_0^{(2)}$. 
The event \{Reject $H_0^{(1)}$, Fail to reject $H_0^{(2)}$ and Reject $H_0^{(3)}$\} is a subset of the event \{Reject $H_0^{(1)}$\}.
Moreover, the probability that we ``Reject $H_0^{(1)}$'' knowing that it is true is at most $\alpha$, thus in this case the family wise error rate is still at most $\alpha$. 
More generally, because testing finer hypotheses is conditional on rejecting a more coarse hypothesis, we don't have to worry about exceeding the family wise error threshold even if we falsely reject a coarser hypothesis. 
In our example, if we falsely reject the $H_0^{(1)}$, then the family wise error rate is $\alpha$ regardless of how many more of the hypotheses we reject. 

Consider another scenario: Suppose $H_0^{(1)}$, $H_0^{(2)}$ and $H_0^{(4)}$ are false. 
Since $H_0^{(4)}$ is false, at most one of $H_0^{(8)}$ and $H_0^{(9)}$ can be true. 
Without loss of generality assume $H_0^{(8)}$ is true and $H_0^{(9)}$ is false. 
Now it is possible that $H_0^{(3)}$ and $H_0^{(5)}$ are also true. 
Therefore, in the worst case the family wise error rate is the probability of falsely rejecting $H_0^{(3)}$, $H_0^{(5)}$ and $H_0^{(8)}$.
Therefore, we test these three hypothesis individually at the $\alpha/3$ significance level. We can swap the roles of $H_0^{(8)}$ and $H_0^{(9)}$ in the above chain of reasoning and deduce that the significance level of $\alpha/3$ must be used for $H_0^{(9)}$ as well. 

From these scenarios we get the following insight: for each hypothesis $H^{(i)}_0$, enumerate which hypotheses can simultaneously be true along with $H^{(i)}_0$. Let $i_1, i_2, ..., i_{k(i)}$ denote the hypotheses that can be true along with $i$, and let $\alpha_{i_1}, \dots ,\alpha_{i_{k(i)}}$ denote the corresponding significance levels. 
Now using the union bound along with the probability of Type-I error (i.e. significance levels) yields the following constraint:
$$
	\alpha_i + \alpha_{i_{1}} + \dots + \alpha_{i_{k(i)}} \leq \alpha.
$$
For each hypothesis we get similar constraints, some of which can coincide. 
Note that there can be multiple constraints for a given hypothesis (as is the case for $H_0^{(3)}$ and $H_0^{(5)}$). 
For instance, the constraint for $H_0^{(8)}$ in our hypothesis tree is $\alpha_8 + \alpha_3 + \alpha_5 \leq \alpha$, and the constraint for $H_0^{(9)}$ in our hypothesis tree is $\alpha_9 + \alpha_3 + \alpha_5 \leq \alpha$. Note that these are also the constraints for $H_0^{(3)}$ and $H_0^{(5)}$. 
The allocation $\alpha_3 = \alpha_5 =  \alpha_8 = \alpha_9 = \alpha_3$ is consistent with these.

For our particular application, the tree structure of Figure~\ref{fig:trt} yields the following set of unique constraints
\begin{align*}
    \alpha_8 + \alpha_3 + \alpha_5 &\leq \alpha\\
    \alpha_9 + \alpha_3 + \alpha_5 &\leq \alpha\\
    \alpha_4 + \alpha_3 &\leq \alpha\\
    \alpha_2 &\leq \alpha\\
    \alpha_1 &\leq \alpha
\end{align*}
The significance levels shown in Figure~\ref{fig:hyp1} satisfy these constraints.
We conjecture that any set of constraints induced by tree-structured hypotheses will have a feasible solution but leave a formal proof for future work.


\end{document}